# Parallel Computation of Finite Element Navier-Stokes codes using MUMPS Solver

Mandhapati P. Raju

[1] Mechanical Engineering, Case Western Reserve University,
Cleveland, Ohio 44106
raju192@gmail.com

**Abstract**
The study deals with the parallelization of 2D and 3D finite element based Navier-Stokes codes using direct solvers. Development of sparse direct solvers using multifrontal solvers has significantly reduced the computational time of direct solution methods. Although limited by its stringent memory requirements, multifrontal solvers can be computationally efficient. First the performance of MUltifrontal Massively Parallel Solver (MUMPS) is evaluated for both 2D and 3D codes in terms of memory requirements and CPU times. The scalability of both Newton and modified Newton algorithms is tested.
*Key words: finite element, MUMPS solver, distributed computing, Newton method.*

## 1. Introduction

Discretization of Navier-Stokes equations involves a large set of non-linear equations, requiring high computing power in terms of speed and memory. The resulting set of weak form (algebraic) equations in such problems may be solved either using a direct solver or an iterative solver. The direct solvers are known for their generality and robustness. The direct solution methods generally involve the use of frontal algorithms [1] in finite element applications. The advent of multifrontal solvers [2] has greatly increased the efficiency of direct solvers for sparse systems. They make full use of the high computer architecture by invoking level 3 Basic Linear Algebra Subprograms (BLAS) library. Thus the memory requirement is greatly reduced and the computing speed greatly enhanced. Multifrontal solvers have been successfully used both in the context of finite volume problems [3-5], finite element problems [6] and in power system simulations [7-9]. The disadvantage of using direct solvers is that the memory size increases much more rapidly than the problem size itself [6]. To circumvent this problem, out-of-core multifrontal solvers [10] have been developed which has the capability of storing the factors on the disk during factorization. Another viable alternative is to use direct solvers in a distributed computing environment.

The system of non-linear equations obtained from the discretization of Navier-Stokes equations is usually solved using a Newton or a Picard algorithm. Newton algorithms are known for their quadratic convergence behavior. When the initial guess is close to the final solution, Newton achieves quadratic convergence. In this paper, only the Newton algorithm is used. In using direct solvers, factorization of the left hand side matrix is the most time consuming step. To avoid factorization during every iteration, a modified Newton is used in which the factorization is done only during the first iteration. The left side matrix evaluated for the first iteration is retained and is not changed during the subsequent iterations. Only the right hand side matrix is updated during each iteration step. The right hand side vector is appropriately modified to give the final converged solution. Since the factorization is done only during the first iteration, the subsequent iterations are extremely cheap. It usually requires more number of iterations to obtain the overall convergence. So there is a tradeoff between the computational time per iteration and the number of iterations to obtain the final convergence. Although the convergence rate is lower compared to the Newton iteration, the savings in computational time per iteration is so high that it can more than compensate the decrease in the convergence rate.

MUMPS [11-13] and SUPERLU [14] are amongst the fastest parallel general sparse direct solvers that are available under public domain software. A detailed description of the various features and algorithms employed in these packages can be found in [15]. MUMPS is found to be much faster compared to SUPERLU, although its scalability is low compared to that of SUPERLU. In this paper, parallelization is achieved using a MUltifrontal Massively Parallel Solver (MUMPS) on a distributed environment using MPI. The linear system of equations is evaluated on different processors corresponding to the local grid assigned to the processor. The right hand side vector is assembled on the host processor and is input to the MUMPS solver. On exit from the MUMPS solver, the solution is assembled centrally on the host processor. This solution is then broadcast to all the processors. In the context of modified Newton algorithm, the LU factors evaluated during the first iteration are reused and the solution of the linear system with the new right hand side vector is solved. The performance of the solver in terms of scalability and memory issues for both two-dimensional and three-dimensional problems are discussed in detail.

## 2. Mathematical Formulation

The governing equations for laminar flow through a two-dimensional rectangular duct are presented below in the non-dimensional form.

$$\frac{\partial u}{\partial x} + \frac{\partial v}{\partial y} = 0, \tag{1}$$

$$\frac{\partial}{\partial x}(u^2) + \frac{\partial}{\partial y}(uv) = -\frac{\partial p}{\partial x} + \frac{\partial}{\partial x}\left(\frac{2}{Re}\frac{\partial u}{\partial x}\right) + \frac{\partial}{\partial y}\left(\frac{1}{Re}\left(\frac{\partial u}{\partial y} + \frac{\partial v}{\partial x}\right)\right), \tag{2}$$

and

$$\frac{\partial}{\partial x}(uv) + \frac{\partial}{\partial y}(v^2) = -\frac{\partial p}{\partial y} + \frac{\partial}{\partial x}\left(\frac{1}{Re}\left(\frac{\partial u}{\partial y} + \frac{\partial v}{\partial x}\right)\right) + \frac{\partial}{\partial y}\left(\frac{2}{Re}\frac{\partial v}{\partial y}\right), \tag{3}$$

where $u, v$ are the $x$ and $y$ components of velocity, $p$ is the pressure. The bulk flow Reynolds number, $Re=\rho U_0 D/\mu$, $U_0$ being the inlet velocity, $\rho$ the density, $L$ the channel length, and $\mu$ is the dynamic viscosity. Velocities are non-dimensionalized with respect to $U_0$, pressure with respect to $\rho U_0^2$.

The boundary conditions are prescribed as follows:
(1) Along the channel inlet:
$$u = 1; \; v = 0. \tag{4}$$

(2) Along the channel exit:
$$p = 0; \; \frac{\partial u}{\partial x} = 0; \; \frac{\partial v}{\partial x} = 0. \tag{5}$$

(3) Along the walls:
$$u = 0; \; v = 0. \tag{6}$$

The governing equations for laminar flow through a three-dimensional rectangular duct are presented below in the non-dimensional form. In three-dimensional calculations, instead of the primitive $u,v,w,p$ formulation, penalty approach is used to reduce the memory requirements.

$$\frac{\partial u}{\partial x} + \frac{\partial v}{\partial y} + \frac{\partial w}{\partial z} = 0, \tag{7}$$

$$\frac{\partial}{\partial x}(u^2) + \frac{\partial}{\partial y}(uv) + \frac{\partial}{\partial z}(uw) = \lambda \frac{\partial}{\partial x}\left(\frac{\partial u}{\partial x} + \frac{\partial v}{\partial y} + \frac{\partial w}{\partial z}\right) + \frac{\partial}{\partial x}\left(\frac{2}{Re}\frac{\partial u}{\partial x}\right) + \frac{\partial}{\partial y}\left(\frac{1}{Re}\left(\frac{\partial u}{\partial y} + \frac{\partial v}{\partial x}\right)\right) + \frac{\partial}{\partial z}\left(\frac{1}{Re}\left(\frac{\partial u}{\partial z} + \frac{\partial w}{\partial x}\right)\right), \tag{8}$$

$$\frac{\partial}{\partial x}(uv) + \frac{\partial}{\partial y}(v^2) + \frac{\partial}{\partial z}(vw) = \lambda \frac{\partial}{\partial y}\left(\frac{\partial u}{\partial x} + \frac{\partial v}{\partial y} + \frac{\partial w}{\partial z}\right) + \frac{\partial}{\partial x}\left(\frac{1}{Re}\left(\frac{\partial u}{\partial y} + \frac{\partial v}{\partial x}\right)\right) + \frac{\partial}{\partial y}\left(\frac{2}{Re}\frac{\partial v}{\partial y}\right) + \frac{\partial}{\partial z}\left(\frac{1}{Re}\left(\frac{\partial v}{\partial z} + \frac{\partial w}{\partial y}\right)\right), \tag{9}$$

and

$$\frac{\partial}{\partial x}(uw) + \frac{\partial}{\partial y}(vw) + \frac{\partial}{\partial z}(w^2) = \lambda \frac{\partial}{\partial z}\left(\frac{\partial u}{\partial x} + \frac{\partial v}{\partial y} + \frac{\partial w}{\partial z}\right) + \frac{\partial}{\partial x}\left(\frac{1}{Re}\left(\frac{\partial u}{\partial z} + \frac{\partial w}{\partial x}\right)\right) + \frac{\partial}{\partial y}\left(\frac{1}{Re}\left(\frac{\partial v}{\partial z} + \frac{\partial w}{\partial y}\right)\right) + \frac{\partial}{\partial z}\left(\frac{2}{Re}\frac{\partial w}{\partial z}\right). \tag{10}$$

where $u, v, w$ are the $x$, $y$ and $z$ components of velocity,. The bulk flow Reynolds number, $Re=\rho U_0 D/\mu$, $U_0$ being the inlet velocity, $\rho$ the density, $L$ the channel length, $\mu$ is the dynamic viscosity and $\lambda$ is the penalty parameter. Velocities are non-dimensionalized with respect to $U_0$.
The boundary conditions are prescribed as follows:
(1) Along the channel inlet:
$$u = 1; \; v = 0; w = 0. \tag{11}$$

(2) Along the channel exit :
$$\frac{\partial u}{\partial x} = 0; \; \frac{\partial v}{\partial x} = 0; \; \frac{\partial w}{\partial x} = 0. \tag{12}$$

(3) Along the walls:
$$u = 0; \; v = 0; \; w = 0. \tag{13}$$

## 4. Newton's Algorithm

The set of non-linear equations obtained by the discretization of Galerkin Finite element formulation is solved using Newton's iterative algorithm.

Let $\underline{X}^{(k)}$ be the available vector of field unknowns for the $k^{\text{th}}$ iteration. Then the update for the $(k+1)^{st}$ iteration is obtained as

$$\underline{X}^{(k+1)} = \underline{X}^{(k)} + \alpha\, \delta \underline{X}^{(k)}, \qquad (14)$$

where $\alpha$ is an under-relaxation factor, and $\delta \underline{X}^{(k)}$ is the correction vector obtained by solving the linearized system

$$[J]\{\delta \underline{X}^{(k)}\} = -\{\underline{R}_X\}^{(k)}. \qquad (15)$$

Here, $[J]$ is the Jacobian matrix,

$$[J] = \frac{\partial \underline{R}_X^{(k)}}{\partial \underline{X}^{(k)}}. \qquad (16)$$

and $\{\underline{R}_X\}^{(k)}$ is the residual vector. Newton's iteration is continued till the infinity norm of the correction vector $\delta \underline{X}^{(k)}$ converges to a prescribed tolerance of $10^{-6}$.

The modified Newton's algorithm employs the jacobian calculated during the first iteration repeatedly for all the successive iterations. The jacobian is not updated. This would reduce the rate of convergence of the Newton algorithm. Since the jacobian is not updated, factorization can be skipped during all the subsequent iterations.

## 5. MUMPS Solver

The most time consuming part is the solution of the set of linear equations. To solve these linear systems, we use a robust parallel solver MUMPS. It employs multifrontal techniques to solve the set of linear equations on parallel computers on a distributed environment using MPI. It relies on Level II and Level III optimized BLAS routines. It requires SCALAPACK and PBLACS routines. In this paper, the vendor optimized INTEL Math Kernal Library is used. The Software is written in Fortran 90 and has a C interface available.

The solution of the set of linear equations takes place in 3 essential steps
(i) Analysis step: MUMPS offers various built in ordering algorithms and interface to external packages such as PORD [16] and METIS [17]. The Matrix is analyzed to determine an ordering and a mapping of the multifrontal computational graph is constructed. This symbolic information is passed on from the host to all the other processors.
(ii) Factorization: Based on the symbolic information, the algorithm tries to construct several dense sub matrices that can be processed in parallel. The numerical factorization is carried out during this step.
(iii) Solution step: Using the right hand side vector, the solution vector is computed using the distributed factors.

All these steps can be called separately or as a combination of each other. This can be exploited to save some computational effort during the solution of subsequent iterations in the solution of a set of nonlinear equations. For example if the structure of the matrix does not change during every iteration, the analysis step can be skipped after evaluating once. Similarly, if the left hand matrix does not change, both the analysis and the factorization steps can be skipped.

## 6. Parallel Implementation

The MUMPS solver is implemented using the MPI library, which makes the code very portable and usable on both, shared and distributed memory parallel computers. The parallelization is done internally in the code. The calling program should also be in a parallel environment to call the code. In the present formulation, each element is assigned to particular processor such the elements are equally (or almost equally) distributed amongst all the processors. The computation of the matrix coefficients and the right hand side vector are done in parallel corresponding to the set of local elements. Evaluation of the Jacobian matrix and the right hand side vector in a parallel environment is crucial for problems, which consume lot of time for the evaluation of matrix coefficients.

During the progress of overall iterations, the different set of linear equations obtained during every iteration is solved by successive calls to the MUMPS. For the modified Newton's algorithm, the left hand matrix remains the same (numerically). So both the analysis and the factorization steps are skipped during the subsequent iterations. Since the factorization is most costly step, it leads to a significant amount of savings in time for the subsequent iterations. The performance of Newton and Modified Newton's method is tested.

While implementing the finite element code in a parallel environment with the MUMPS code, the element matrix entries are calculated locally on each of the processors. Although the facility for element matrix input is available, only the option of centralized element entry is available in the current versions of MUMPS solver. To facilitate distributed matrix input (necessary for improving the parallel efficiency), the local element entries are converted into sparse matrix triplet entries in coordinate format and are input in a distributed fashion (using ICNTL(5) = 0 and ICNTL(18) = 3). There will be lot of duplicate entries due to contribution of all the neighboring elements at a given grid point. MUMPS solver automatically sums up all the duplicate entries. Different ordering can be chosen by using different values for ICNTL(7). The different ordering options that are available within MUMPS solver are (i) Approximate minimum degree (AMD), (ii)

Approximate minimum fill (AMF), (iii) PORD, (iv) METIS, (v) Approximate Minimum degree with automatic quasi-dense row detection (QAMD).

## 7. Performance of MUMPS Solver

Preliminary experiments have been done to study the effect of different ordering routines on the performance of MUMPS solver both in terms of memory and computational time. Table 1 shows the performance of different ordering routines for two-dimensional codes. Table 1 shows the comparison of different ordering routines for a 300x300 mesh using 12 processors. Results indicate the both PORD and METIS perform well in minimizing the computational time requirements but METIS performs well in terms of memory requirements. Based on this observation, METIS ordering is used for all subsequent calculations. Table 2 shows the performance of Newton's and modified Newton's method using MUMPS solver for a two-dimensional channel flow problem. Results indicate that modified Newton's method performs better than Newton's method. This is due to the fact that in modified Newton's method, factorization is done only during the first iteration. During the subsequent iterations, factorization is skipped and only the solution step is performed. This decreases the convergence rate, thereby increasing the number of iterations to obtain convergence. However, the solution step being computationally inexpensive compared to the factorization step, the overall solution time is less compared to the Newton's step. However it should be noted that the memory requirement for the Newton and modified Newton's method are the same. Table 2 shows that the memory requirement does not vary linearly with the number of processors. It behaves erratically. Table 2 also shows that the MUMPS solver does not scale so well. Using 20 processors, the computational time is approximately halved compared to the computational time using 2 processors. It does not scale much beyond 6 processors. The scalability of MUMPS solver for two-dimensional problems is observed to be poor.

Table 3 shows the performance of MUMPS solver using different ordering routines for a three dimensional channel flow problem. The table shows that METIS ordering is a better choice both in terms of computational speed and memory requirement. Hence METIS is used for all the subsequent computations of three dimensional problems. For a 100x20x20 mesh, memory was not sufficient to run on 2 processors. The table shows that the scalability of MUMPS solver for three-dimensional problems is better than that for the two-dimensional problems. When the number of processors increased from 4 to 20, the computational time of Newton's method has reduced to a factor of 3.6 approximately, while that of modified Newton's method has reduced to a factor of 3 approximately. The maximum memory requirement for a single processor has reduced to a factor of 4. The use of MUMPS solver for three dimensional problems seems to be promising. However, memory requirement is a serious limitation for solving three dimensional problems using direct solvers.

## 8. Conclusions

Finite element based Navier-Stokes codes are parallelized using MUMPS solver. Both Newton and modified Newton's algorithms are used. It is observed that modified Newton's method leads to savings in computational time compared to the Newton's algorithm. It is also observed that METIS ordering enhances the performance of MUMPS solver both for two-dimensional and three-dimensional problems. MUMPS solver does not scale well for two-dimensional problems but it scales better for three dimensional problems.

Table 1: Comparison of the performance of different orderings in MUMPS solver for a 300x300 mesh on 12 processors

| Ordering | CPU time/ iteration (sec) | Memory (MB) | |
|---|---|---|---|
| | | avg | max |
| PORD | 16.6 | 1525 | 1640 |
| METIS | 5.1 | 1560 | 1820 |
| AMD | 5.5 | 1586 | 1923 |
| AMF | 6.7 | 1592 | 2000 |
| QAMD | 5.3 | 1603 | 2056 |

Table 2: Comparison of the performance of Newton and Modified Newton's methods using MUMPS solver for a 200x200 mesh

| # of processors | Time to solve (Seconds) | | Memory Requirements (MB) | | Ordering |
|---|---|---|---|---|---|
| | Newton | Modified Newton | max memory on one processor | Total memory | |
| 2 | 35.7 | 27.4 | 412 | 806 | Metis |
| 4 | 23 | 18.4 | 259 | 977 | Metis |
| 6 | 21 | 16.8 | 227 | 1082 | Metis |
| 8 | 20 | 15.3 | 178 | 1209 | Metis |
| 10 | 19.2 | 14.8 | 159 | 1394 | Metis |
| 12 | 18.6 | 14.6 | 157 | 1700 | Metis |
| 14 | 19.1 | 15 | 171 | 2039 | Metis |
| 16 | 18.4 | 13 | 156 | 2352 | Metis |
| 18 | 18.3 | 13 | 145 | 2534 | Metis |
| 20 | 18 | 12 | 141 | 2675 | Metis |

Table 3: Comparison of the performance of different orderings in MUMPS solver for a 100x20x20 mesh on 12 processors

IJCSI

| Ordering | CPU time/iteration (sec) | Memory (MB) | |
|---|---|---|---|
| | | Avg | max |
| PORD | 41 | 1385 | 1612 |
| METIS | 38 | 1286 | 1400 |
| AMD | 105 | 2296 | 2496 |
| AMF | 93 | 1246 | 1425 |
| QAMD | 102 | 2296 | 2593 |

Table 4: Comparison of the performance of Newton and Modified Newton's methods using MUMPS solver for a 100x20x20 mesh

| # of processors | Time to solve (Seconds) | | Memory requirements (GB) | | Ordering |
|---|---|---|---|---|---|
| | Newton | Modified Newton | max memory on one processor | Total memory | |
| 4 | 184 | 89 | 2 | 7.3 | metis |
| 6 | 147 | 77.2 | 1.6 | 8.6 | metis |
| 8 | 112 | 61.6 | 1 | 7.5 | metis |
| 10 | 91 | 53.4 | 1.5 | 13.9 | metis |
| 12 | 99 | 42.5 | 1.4 | 15.4 | metis |
| 14 | 68 | 37.2 | 1.45 | 17 | metis |
| 16 | 58 | 37 | 0.7 | 9.6 | metis |
| 18 | 52 | 34.8 | 0.63 | 10.2 | metis |
| 20 | 50 | 31.1 | 0.56 | 9.9 | metis |

**Acknowledgments**

Author would like to thank the Ohio Supercomputing Centre for proving the computing facility.

**References**

[1] B.M. Irons, "A frontal solution scheme for finite element analysis," **Numer. Meth. Engg,** Vol. 2, 1970, pp. 5-32.
[2] T. A. Davis, and I. S. Duff, "A combined unifrontal/multifrontal method for unsymmetric sparse matrices," **ACM Trans. Math. Softw.**, Vol 25, No 1, 1997, pp. 1–19.
[3] M. P. Raju, and J. S. T'ien, "Development of Direct Multifrontal Solvers for Combustion Problems," **Numerical Heat Transfer, Part B,** Vol. 53, 2008, pp. 1-17.
[4] M. P. Raju, and J. S. T'ien, "Modelling of Candle Wick Burning with a Self-trimmed Wick," **Comb. Theory Modell.,** Vol. 12, No. 2, 2008, pp. 367-388.
[5] M. P. Raju, and J. S. T'ien, "Two-phase flow inside an externally heated axisymmetric porous wick," **Journal of Porous Media,** Vol. 11, No. 8, 2008, pp. 701-718.
[6] P. K. Gupta and K. V. Pagalthivarthi, "Application of Multifrontal and GMRES Solvers for Multisize Particulate Flow in Rotating Channels," **Prog. Comput Fluid Dynam.,** Vol. 7, 2007, pp. 323–336.
[7] S. Khaitan, J. McCalley, Q. Chen, "Multifrontal solver for online power system time-domain simulation," **IEEE Transactions on Power Systems**, Vol. 23, No. 4, 2008, pp. 1727–1737.
[8] S. Khaitan, C. Fu, J. D. McCalley, "Fast parallelized algorithms for online extended-term dynamic cascading analysis," **PSCE**, 2009, pp. 1–7.
[9] J. McCalley, S. Khaitan, "Risk of Cascading outages", Final Report, PSrec Report, S-26, August 2007. http://www.pserc.org/docsa/Executive_ Summary_Dobson_ McCalley_Cascading_Outage_ S-2626_PSERC_ Final_ Report.pdf.
[10] J. A. Scott, Numerical Analysis Group Progress Report, RAL-TR-2008-001.
[11] P. R. Amestoy, I. S. Duff and J.-Y. L'Excellent, "Multifrontal parallel distributed symmetric and unsymmetric solvers," **Comput. Methods in Appl. Mech. Eng.,** Vol. 184, 2000, pp. 501-520.
[12] P. R. Amestoy, I. S. Duff, J. Koster and J.-Y. L'Excellent, "A fully asynchronous multifrontal solver using distributed dynamic scheduling," **SIAM Journal of Matrix Analysis and Applications,** Vol. 23, No. 1, 2001, pp 15-41.
[13] P. R. Amestoy, A. Guermouche, J.-Y. L'Excellent and S. Pralet, "Hybrid scheduling for the parallel solution of linear systems," **Parallel Computing,** Vol. 32, No. 2, 2006, pp. 136-156.
[14] S. L. Xiaoye and W. D. James, "A Scalable Distributed-Memory Sparse Direct Solver for Unsymmetric Linear Systems," **ACM Trans. Mathematical Software,** Vol. 29, No. 2, 2003, pp. 110-140.
[15] A. Gupta, "Recent advances in direct methods for solving unsymmetric sparse systems of linear equations," **ACM transaction in Mathematical Software,** Vol. 28, No. 3, 2002, pp. 301-324.
[16] J. Schulze, "Towards a tighter coupling of bottom-up and top-down sparse Matrix ordering methods," **BIT Numerical Mathematics,** Vol. 41, No. 4, 2001, pp. 800-841.
[17] G. Karypis, and V. Kumar, "A Fast and High Quality Multilevel Scheme for Partitioning Irregular Graphs," **SIAM J Scientific Computing,** Vol. 20, 1999, pp. 359-392.

**Mandhapati P. Raju**

Mandhapati P. Raju completed his MS (2002-2004) and Ph.D (2004-2006) in Mechanical Engineering department at Case Western Reserve University, Cleveland, OH. Later he worked as Postdoctoral fellow in Case Western Reserve University during 2006-2008. Later he worked at Caterpillar Champaign simulation centre as a CFD analyst. Currently he is working as a Post Doctoral fellow in General Motors Inc. His research interests are combustion, porous media flows, multifrontal solvers, fuel cell and hydrogen storage. This work was done during his presence in Case Western Reserve University. He has published 5 journal papers in reputed international journals.